\documentclass[12pt,halfline,a4paper,]{ouparticle}

\usepackage{listings}
\usepackage{hyperref}
\usepackage{graphicx}
\usepackage{listings}
\usepackage{xcolor}
\usepackage{fancyvrb}
\usepackage{framed}

\hypersetup{breaklinks=true,
            pdfauthor={},
            pdftitle={Multimodal LLM, Geospatial AI, and Urban Policy: Estimating Redlining's Legacy effects using Street-View Imagery}
            }

\usepackage[T1]{fontenc}
\usepackage{lmodern} 

\usepackage{amssymb, amsmath, geometry, amsfonts, verbatim, endnotes, setspace}

\IfFileExists{upquote.sty}{\usepackage{upquote}}{}

\makeatletter
\def\Newlabel#1#2#3{\expandafter\gdef\csname #1@#2\endcsname{#3}}

\def\Ref#1#2{\@ifundefined{#1@#2}{???}{\csname #1@#2\endcsname}}

\newcommand*\ifcounter[1]{%
  \ifcsname c@#1\endcsname
    \expandafter\@firstoftwo
  \else
    \expandafter\@secondoftwo
  \fi
}

\usepackage{setspace}
\usepackage{booktabs}
\usepackage{float}
\usepackage{caption}
\usepackage{cite}
\usepackage{array,arydshln}
\usepackage{xcolor}
\usepackage{threeparttable}
\usepackage[utf8x]{inputenc}
\usepackage{arydshln}
\usepackage{soul}
\usepackage{lineno}
\DeclareUnicodeCharacter{2212}{-}

\begin{document}

\title{From Pixels to Urban Policy-Intelligence: Recovering Legacy Effects of Redlining with a Multimodal LLM}

\author{%
%
%
%
%
\name{%
  Anthony Howell\(^{\dagger,*}\),
  Nancy Wu\(^{\ddagger}\),
  Sharmistha Bagchi\textendash Sen\(^{\dagger}\),
  Yushim Kim\(^{\dagger}\),
  Chayn Sun\(^{\S}\)
}
}
\abstract{This paper shows how a multimodal large language model (MLLM) can expand urban measurement capacity and support tracking of place-based policy interventions. Using a structured, reason-then-estimate pipeline on street-view imagery, GPT-4o infers neighborhood poverty and tree canopy, which we embed in a quasi-experimental design evaluating the legacy of 1930s redlining. GPT-4o recovers the expected adverse socio-environmental legacy effects of redlining, with estimates statistically indistinguishable from authoritative sources, and it outperforms a conventional pixel-based segmentation baseline—consistent with the idea that holistic scene reasoning extracts higher-order information beyond object counts alone.  These results position MLLMs as policy-grade instruments for neighborhood measurement and motivate broader validation across policy--evaluation settings.}

\date{\today}


\maketitle

\noindent\small \(^{1}\)Arizona State University

\noindent\small \(^{2}\)University at Albany - State University of New York

\noindent\small \(^{3}\)Royal Melbourne Institute of Technology

\noindent\small \(^{*}\)Corresponding author:
\texttt{Anthony.Howell@asu.edu}

\newpage

\section{Introduction}

Multimodal large language models (MLLMs) are emerging as general-purpose systems which, alongside urban foundation models, offer wide-ranging potential for downstream city applications— from transportation and urban planning to environmental monitoring, energy management, and public safety \cite{zhang2024ufm,mai2024fmgeoai,resch2025fmhealth,roberts2024charting,ji2025geospatialreasoning,yuan2025omnigeo}. Through holistic scene reasoning, MLLMs process both visual and textual inputs and could be relied on to produce structured urban indicators alongside natural-language rationales that explain the visual cues behind those predictions \cite{achiam2023gpt4,openai2024gpt4o}. Although MLLMs are being explored for geospatial reasoning and urban intelligence functions \cite{li2024streetviewllm}, evidence on their utility for spatial descriptive diagnostics—and especially as a policy-evaluation instrument—remains limited.

Consistent with recent calls in sustainability science to leverage recent advances in AI to extend measurement capacity for SDG-relevant indicators \cite{gohr2025artificial}, this paper examines whether MLLM-derived neighborhood measures from Google Street View (GSV) imagery can recover intervention--outcome relationships comparable to those obtained with authoritative data under a shared identification design. Our policy context is the long-run relationship between redlining—a discriminatory housing policy institutionalized in the 1930s by the Home Owners’ Loan Corporation (HOLC) \cite{hillier_redlining_2003}—and present-day outcomes, with the Phoenix Metropolitan Statistical Area (MSA) as a testbed. We focus on poverty and tree canopy, two widely used indicators of sustainable urban development (e.g., SDGs 1 and 11) \cite{fan2023urban,ye2019measuring} that are visually salient in GSV imagery and directly relevant to the legacy of redlining \cite{appel2016pockets,locke2021residential,aaronson_effects_2021,nardone_redlines_2021,salazar_long_2024,gibbons_evaluating_2025}.

Redlining systematically denied mortgage credit to neighborhoods with higher proportions of minority residents. These areas—typically denser, older sections of the urban core—were marked as “hazardous” (HOLC-D; red) on appraisal maps and became subject to long-term disinvestment, leading to persistent socioeconomic disadvantage and degraded environmental quality \cite{appel2016pockets,locke2021residential}. As capital flowed instead to suburban, predominantly white communities, redlined neighborhoods faced restricted access to homeownership and wealth accumulation, deterioration of housing stock, and reduced public and private investment in infrastructure, parks, and greenspace \cite{kruse2006new,grove2018legacy}.

To capture these linked dimensions from GSV imagery, we apply GPT-4o—an MLLM with native vision capabilities—in a structured reason-then-estimate pipeline inspired by chain-of-thought prompting \cite{wei2022chain}. In the first stage, the model produces auditable, schema-based descriptions of visible street-view features (SVFs), e.g., housing type, facade condition, infrastructure quality, and vegetation. In the second stage, those structured cues are fed to a follow-up prompt that maps them to standardized probabilistic indicators (e.g., share of households below poverty thresholds, tree-canopy coverage). Predictions are stabilized via five-round self-consistency \cite{wang2022self} (averaged across five passes per image), and then aggregated across images to the census block group (CBG) level to produce neighborhood-scale proxies.

For comparison, we also derive geospatial predictions using a semantic segmentation model applied to the same imagery. This approach reflects a widely used GeoAI technique for extracting socioeconomic and environmental indicators from street-level imagery \cite{zhang2024urban,li_predicting_2022,zhao_sensing_2023,wang_investigating_2024,suel_multimodal_2021}. In semantic segmentation, computer vision models classify each pixel into predefined categories (e.g., tree canopy, sidewalks, building façades). These pixel-level classifications are then aggregated into neighborhood-level features—such as greenspace availability or signs of socioeconomic deprivation—that serve as proxies for local conditions \cite{gong2018mapping,zou2021detecting,vallebueno2023measuring}.

In total, we process more than 25,000 GSV images, aggregate each set of geospatial predictions to 1,145 CBGs, and merge them with authoritative measures of poverty and tree canopy from the 2023 American Community Survey (ACS) and Google’s Environmental Insights Explorer (GEIE), respectively. Under a shared quasi-experimental framework, we estimate treatment effects separately for each measurement approach using covariate-adjusted regressions with zip code fixed effects (absorbing jurisdiction-level, time-invariant factors) and a spatial autoregressive (SAR) specification (capturing spatial dependence and local spillovers). To formally assess between-approach equivalence, we pool all outcomes in a stacked-regression framework with CBG-level nonparametric bootstrapping, testing whether differences in estimated treatment effects relative to the authoritative benchmark are statistically distinguishable from zero.

The main results show that GPT-4o reproduces patterns that are evident in authoritative data: compared to non-redlined areas, historically redlined neighborhoods continue to exhibit higher socioeconomic disadvantage via higher poverty and lower environmental quality via lower tree canopy.  These findings not only align with ACS and GEIE benchmarks for Phoenix but also mirror well-documented patterns across the U.S. \cite{appel2016pockets,locke2021residential}. 

MLLM-derived poverty measures align more closely with ACS’s survey-based estimates than those from segmentation. One explanation is that holistic scene reasoning integrates higher-order contextual cues and object relations (e.g., infrastructural neglect, maintenance) that pixel-based semantic segmentation may miss. Consistent with this idea, GPT-4o explains more variance on average—with the largest gains at the distributional extremes ( high-poverty, high-canopy neighborhoods)—and a simple fusion of segmentation outputs with GPT-4o yields only marginal improvements. Notably, GPT-4o’s predictions explain variance comparable to a standard covariate set (e.g., demographics, human capital), and combining GPT-4o with these covariates delivers the best overall fit.

Our study contributes to urban visual intelligence and advances toward urban policy-intelligence by: (i) evaluating an MLLM within a policy-evaluation framework that holds identification constant across outcomes and data sources; (ii) showing that GPT-4o–derived measures can reproduce treatment-effect estimates obtained from authoritative data; and (iii) demonstrating that MLLMs can outperform conventional segmentation at capturing contextually important variation. The pattern of results we observe aligns with the extant literature—historically redlined neighborhoods face persistently worse socio-environmental conditions and greater urban sustainability challenges—leading to the same policy conclusions as analyses based on authoritative sources \cite{aaronson_effects_2021,nardone_redlines_2021,salazar_long_2024,gibbons_evaluating_2025}. While generalizability across cities, imaging conditions, and other types of interventions merits further testing, this paper outlines a practical pathway to urban policy-intelligence.

\section{Main Results}\label{main-results}

\subsection{MLLM Inference for Urban Policy-Intelligence}\label{mllm-inference-for-quasi-experimental-policy-analysis}

We evaluate how historical redlining continues to shape contemporary socio-environmental outcomes and how policy conclusions compare across three measurement approaches: (i) authoritative benchmarks (ACS poverty; GEIE canopy), (ii) MLLM-derived estimates from GSV imagery (GSV–GPT-4o), and (iii) a pixel-based semantic-segmentation model (GSV–ResNet). Fig.~\ref{fig:fig1_combinedmap} maps ACS poverty, GEIE canopy, and a composite Sustainability Index (SI $\equiv Z(\text{canopy}) - Z(\text{poverty})$) that provides a snapshot of each neighborhood’s socio-environmental profile; reports variable means by HOLC grouping (See also Table~\ref{tab:summary_by_holc}); and illustrates two pipelines applied to the same GSV imagery: a structured GPT-4o prompt set (Appendix~\ref{prompts}) and a ResNet-based segmentation model.

We estimate treatment effects of redlining, aiming to determine if the direction and magnitude of the effects are similar when using GPT-4o-derived outcome measures of poverty and tree canopy compared to authoritative measures from the ACS and GEIE. To assess robustness, we benchmark HOLC--D (redlined) neighborhoods against two reference groups drawn from the HOLC system: an {Ideal} group (A) and a {Stable/Declining} group (B/C). Because rated A and B/C CBGs are scarce in Phoenix, each benchmark is augmented with closely matched, non-rated CBGs (Pseudo-A; Pseudo-B/C) to increase precision while preserving group comparability. We expect the estimated redlining effects to be larger against the Ideal group compared to the Stable/Declining group.

To assess how stable the estimated redlining effects are across model specifications, we estimate—separately for each measurement approach—a sequence of variants of Eq.\eqref{eq:core-m}: We start with an unadjusted model (raw redlined vs. non-redlined difference), then add a covariate-adjusted model with standard neighborhood, demographic, and human-capital controls. Next, we fit a zip code fixed-effects model to absorb unobserved heterogeneity at the jurisdiction level that are time-invariant, such as amenities, service provision, zoning history, and tax base. Finally, we estimate a spatial-lag (SAR) model to account for spatial dependence and local spillovers. 

The main results show that as covariates, zip code fixed effects, and spatial structure are sequentially added, estimated redlining effects attenuate in a similar pattern across all three measurement approaches and for both reference groups (Fig.~\ref{fig:FigPolEvalresults}; Tables~\ref{HOLC coefficient estimates – Poverty}–\ref{HOLC coefficient estimates – Sustainability}). Both GSV–based approaches (GPT-4o; ResNet) reproduce the specification sensitivity observed with the authoritative benchmarks (ACS/GEIE). Visual comparisons of point estimates and 95\% confidence intervals show comparable magnitudes for canopy and for the composite Sustainability Index across specifications and comparison groups.

Estimated treatment effects on the Sustainability Index, for instance, tend to concentrate around $-0.30$ SD with overlapping 95\% CIs across approaches, indicating that, regardless of measurement approach, historically redlined neighborhoods continue to exhibit persistently worse socio-environmental conditions. The notable exception is poverty under GSV–ResNet, where effects are noticeably smaller than the authoritative benchmark in two of the four specifications. In the preferred SAR specification, ACS implies a redlining–poverty contrast of $0.35$ SD (SE $= 0.085$), whereas GSV–ResNet yields $0.21$ SD (SE $= 0.103$), consistent with under-detection when relying primarily on pixel counts.

Policy implications depend on the outcome. For canopy and the Sustainability Index, GSV–based estimates lead to the same conclusions as the authoritative benchmarks. For poverty, conclusions align when using GSV–GPT-4o, but the pixel-based segmentation baseline attenuates the estimated effect and can fail to detect disparities, potentially leading to different policy inferences.

\subsubsection{Equivalence Testing via Stacked Regression}
\label{sec:stacked-equivalence}

While visual 95\% CI comparisons are suggestive, they do not test equivalence directly. We therefore use a stacked design that facilitates formal tests of equivalence. All approaches are combined into a unified framework with (i) a single CBG sample, (ii) a uniform specification, and (iii) identically standardized outcomes. We perform 500 nonparametric cluster bootstraps at the CBG level to derive empirical distributions of treatment effects and corresponding 95\% CIs. By including methodological approach indicators and their interactions with redlining (Eq.~\ref{eq:stacked}), the stacked regression directly parameterizes between-method differences in the estimated redlining effects. This setup enables formal hypothesis tests of whether treatment effects derived from GSV measures are statistically different from those obtained from authoritative ACS/GEIE benchmarks. 

The violin plots in Fig.~\ref{fig:Bootstrap_Interaction_Effects} display the empirical bootstrap distributions of total estimated redlining effects $\delta^{(m)}$ for each measurement approach $m$ under both the zip code FE and SAR specifications. The bootstrap distributions confirm that GPT-4o produces redlining effects statistically indistinguishable from the authoritative ACS/GEIE benchmark across all outcomes and specifications. By contrast, ResNet aligns with ACS/GEIE for canopy and the Sustainability Index but systematically underestimates poverty effects, consistent with under-detection when relying on pixel-level counts.

The ACS-based baseline estimate of the redlining effect, for instance, is 0.64 SD (95\% CI: [0.56, 0.72]). The GSV--GPT-4o estimate is similar at 0.58 SD (95\% CI: [0.50, 0.64]); here the bootstrap interaction test fails to reject $H_0:\theta_{\text{GPT-4o}}=0$. By contrast, the GSV--ResNet estimate is smaller and less precise (0.23 SD; 95\% CI: [0.09, 0.35]); here the bootstrap test rejects $H_0:\theta_{\text{ResNet}}=0$, indicating that the ResNet-derived treatment effect is significantly smaller than the authoritative ACS benchmark.

\section{Explanations: Assessing Model Performance}\label{benchmarking-and-validating-mllm-geospatial-predictions} 

\subsection{Comparing GPT-4o and ResNet Model Fit}
Having shown that GPT-4o reproduces ACS-based poverty effects and broadly aligns with GEIE-based canopy effects—while ResNet tends to attenuate poverty—we examine the extent to which these methods differ in capturing information relevant to the targeted indicators. Treating the authoritative ACS poverty, GEIE canopy, and sustainability index (SI) as outcomes, we compare bootstrapped $R^2$ (500 resamples) across five specifications: (i) GPT-4o only, (ii) ResNet only, (iii) demographics only, (iv) GPT-4o + ResNet, and (v) GPT-4o + demographics (Fig.~\ref{fig:rsquared_analysis}, Panel A; Tables~\ref{tab:poverty_regression}–\ref{tab:sustainability_regression}). 

For poverty, GPT-4o alone attains $R^2 = 0.60$ (95\% CI: [0.54, 0.65]), substantially above ResNet with $R^2 = 0.29$ (95\% CI: [0.24, 0.33]). The ResNet value is within the range reported by \cite{fan2023urban} across cities (0.10–0.50). Demographics alone reach $R^2 = 0.56$ (95\% CI: [0.52, 0.60]), making GPT-4o and controls comparable in fit (overlapping 95\% CIs). Adding ResNet to GPT-4o yields no gain, whereas adding demographics to GPT-4o increases fit to $R^2 = 0.69$ (95\% CI: [0.65, 0.73]), indicating that GPT-4o contributes visual information beyond what is captured by traditional covariates. 

For tree canopy, GPT-4o explains $R^2 = 0.40$ of the variance, clearly above ResNet ($R^2 = 0.16$). Neither ResNet nor demographics materially improves on GPT-4o, suggesting GPT-4o already captures salient visual features for canopy. For the sustainability index, GPT-4o leads with $R^2 = 0.61$, roughly double ResNet; adding demographics yields a modest increase to $R^2 = 0.66$, underscoring complementarity between image-derived and conventional information. 

\subsection{Heterogeneity in Model Fit Comparisons} \label{model-fit-comparisons-hetero} 

To check for potential heterogeneity by neighborhood conditions, Panel~B of Fig.~\ref{fig:rsquared_analysis} plots, for selected outcome quantiles $\tau\in\{0.10,0.25,0.50,0.75,0.90\}$, the bootstrapped pseudo-$R^2$ from $\tau$--quantile regressions using GPT--4o predictions (vertical axis) against those using ResNet predictions (horizontal axis). Points above the $45^\circ$ line indicate GPT--4o outperforms ResNet at that quantile. GPT--4o exceeds ResNet at every $\tau$, with the largest gaps at upper quantiles (e.g., $\tau\in\{0.75,0.90\}$), indicating stronger performance in higher--poverty and higher--canopy neighborhoods.

For the sustainability index, GPT-4o’s advantage is positive but comparatively flat across quantiles, consistent with variance compression from combining poverty and canopy. This pattern is consistent with the capacity to detect both extremes of the urban landscape: (i) higher-canopy neighborhoods with mature trees and affluent infrastructure; and (ii) higher-poverty areas marked by barren lots and housing decay—so explanatory power remains stable and high across the index distribution.

Across OLS and quantile analyses, GPT‑4o explains more variation in the authoritative outcomes than ResNet, with the largest margins where policy inference often needs the most reliable signal (high‑poverty settings). This pattern is consistent with GPT‑4o’s holistic scene reasoning capturing contextual indicators—maintenance, infrastructure quality, housing conditions—that are not fully reflected in object counts from pixel‑based semantic segmentation.

\section{Discussion}\label{discussion}

This study evaluates whether a leading MLLM can recover the same intervention--outcome conclusions as those obtained with authoritative data. We focus on the legacy effects of redlining on present--day poverty and tree canopy. Using a chained, reason-then-estimate pipeline on GSV imagery, we produce neighborhood--scale proxies for poverty and canopy that closely track ACS poverty and GEIE canopy patterns across the Phoenix MSA.

The main results show that a carefully prompted MLLM (GSV--GPT-4o) recovers redlining treatment effects that align with authoritative benchmarks and achieves higher explanatory power (\(R^2\)) when predicting ACS/GEIE outcomes than a semantic--segmentation baseline (GSV--ResNet), especially for poverty. While pixel-level segmentation excels at counting discrete objects (e.g., pixels labeled ``tree''), poverty is partly encoded in higher--order context---maintenance, disrepair, infrastructure neglect, and housing form---that no single object class captures. By explicitly eliciting those contextual cues before mapping them to probabilistic indicators, GPT--4o aligns more closely with ACS poverty rates, with the largest gains in high--poverty neighborhoods where policy needs are greatest. For canopy, where the target is effectively an object share, the two approaches perform similarly in practice.

The MLLM prompting strategy returns concise, structured evidence for each estimate and requires no task-specific training, lowering the fixed costs of conventional segmentation stacks. In the near term, this enables rapid measurement of key indicators (including in data--poor settings or between statistical releases), use as outcomes in standard descriptive or quasi-experimental evaluations, and transparent diagnostics that support field verification and guide remedial actions (e.g., infrastructure maintenance, greening). Because the MLLM pipeline is scalable, interpretable, and easy to deploy, it offers a practical pathway toward {urban policy-intelligence}: turning ubiquitous street-view images into credible inputs for evidence--based governance.

\section{Data and Methods}\label{data-and-methods}

\subsection{Study Area}\label{study-area-sample}
To maximize internal validity and manage token costs in MLLM prompting, we restrict analysis to a single metropolitan area and a common observation year. We focus on the Phoenix MSA, which provides an ideal testbed for our application. As the hottest major city in the U.S., Phoenix records some of the nation’s highest rates of heat-related illness and mortality—risks that disproportionately affect socioeconomically disadvantaged populations, particularly those residing in formerly redlined neighborhoods more exposed to health and climate hazards \cite{gee2008multilevel,salazar2024long}. The combination of extreme heat and uneven greenspace provision makes the joint distribution of poverty and tree canopy a particularly salient urban policy problem.

\subsection{Authoritative Benchmark Data Sources}\label{authoritative-benchmarks}
Authoritative outcomes are (i) the share of residents below the federal poverty line at the CBG level from the 2023 ACS five‑year file, and (ii) tree canopy coverage from GEIE for 2023. We also construct a composite Sustainability Index, $\mathrm{SI} \equiv Z(\text{tree canopy}) - Z(\text{poverty})$, to summarize neighborhood socio–environmental conditions on a single standardized scale (higher SI indicates greener and less impoverished areas).

\subsection{Street View Acquisition}\label{gsv-collection}
We queried GSV for each CBG and retrieved images in four cardinal directions (0$^\circ$, 90$^\circ$, 180$^\circ$, 270$^\circ$). To promote seasonal consistency, we filtered for captures in 2023 rending an initial retrieval of  24{,}032. We removed unusable tiles (e.g., blank/placeholder frames, “image not available”) using simple image‑quality heuristics and metadata checks; CBGs with fewer than 10 valid images were excluded. The final dataset contains 19{,}480 valid images across 1{,}145 CBGs.

\subsection{Geospatial Prediction Pipelines}\label{prediction-pipelines}
We generate GSV–based outcomes using two complementary pipelines that operate on the same imagery: (i) a MLLM pipeline (GSV--GPT‑4o) and (ii) a pixel‑level semantic segmentation pipeline (GSV--ResNet).  For both pipelines, we compute image‑level predictions  and take the average across all valid images weighted by the number of tiles in a CBG, and standardize all outcomes across CBGs. We merge the GSV–based CBG outcomes with ACS poverty and GEIE canopy to create a common analysis file used in Section~\ref{analytical-approach-to-evaluating-mllm-predictions-for-policy-inference}. We examine correlations between GSV--GPT‑4o, GSV--ResNet, and authoritative outcomes for poverty, canopy, and SI (see Fig.~\ref{fig:fig1_combinedmap} and Fig.~\ref{fig:correlation_heatmap_ordered}). These diagnostics confirm that both GSV pipelines produce neighborhood‑level patterns broadly aligned with ACS/GEIE, motivating the formal equivalence tests reported later.

\subsubsection{MLLM (GPT‑4o)}\label{mllm-gpt4o}
We implement a chained prompt sequence that first elicits structured, auditable descriptions of visible SVFs  and then maps those cues to probabilistic indicator values on a [0,1] scale (e.g., local‑scene proxy for the share of households below poverty thresholds; tree canopy share). Calibration bands and JSON‑only outputs enforce consistent units and discourage placeholder responses; person‑level inferences are disallowed. To stabilize predictions, we run five independent rounds per image and average the outputs (self‑consistency). Image‑level predictions are then averaged within CBGs, weighted by the number of tile imagers, to obtain neighborhood‑scale proxies aligned with the ACS/GEIE units (see Appendix~\ref{prompts} for the exact prompts).

\subsubsection{Semantic Segmentation (ResNet)}\label{segmentation-resnet}
As a conventional GeoAI benchmark, we apply a ResNet50dilated–PPM\_deepsup model pretrained on ADE20K \cite{zhou2019semantic} to classify pixels into urban semantic categories. Following established practice \cite{naik2017computer,fan2023urban}, we aggregate the original classes into a compact set of SVFs relevant to outdoor form and infrastructure (e.g., building façades/openings, roads/sidewalks, street furniture, grass/shrub, tree canopy, sky). For each image we compute the pixel share of each SVF across the four directions; canopy coverage is proxied by the “tree canopy” share. To avoid sensitive content, person‑level categories are not used as proxies. All SVF shares are averaged to the CBG level, weighted by the number of tiles, to produce neighborhood predictions.

\subsection{Empirical Framework}\label{analytical-approach-to-evaluating-mllm-predictions-for-policy-inference}

\subsubsection*{Defining Treatment and Comparison Groups}

\begin{figure}[!htbp]

\begin{center}\includegraphics[width=1\linewidth]{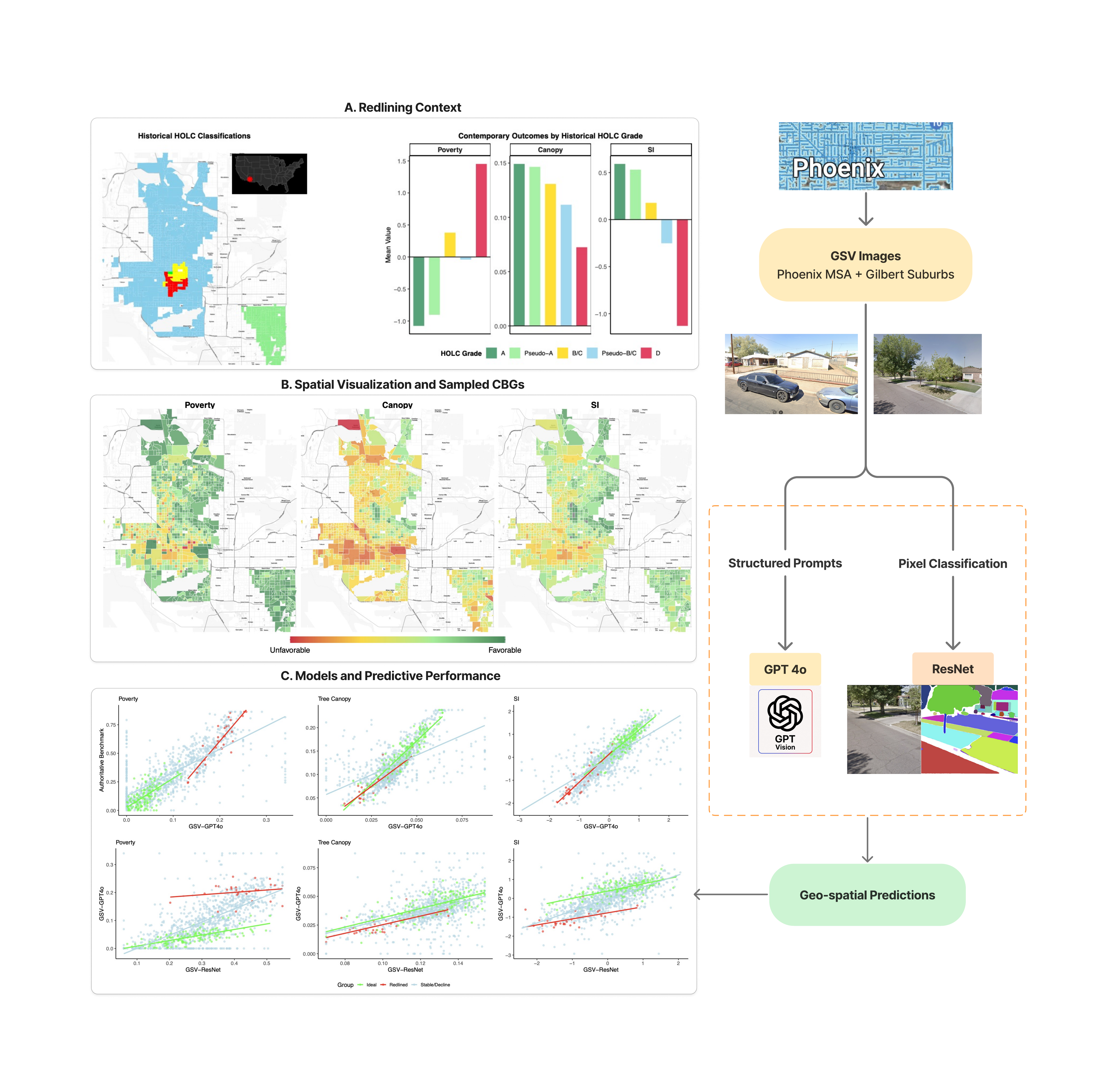} \end{center}

\caption{Spatial Distribution of main variables across Maricopa County, Arizona}
\label{fig:fig1_combinedmap}
\vspace{0.05in}
\scriptsize\textit{Notes:}
{Panel A (redlining context):} Sample sites showing HOLC--rating map and group means for ACS poverty, GEIE canopy, and a composite sustainability index  (SI = $z$(poverty) − $z$(canopy)) that captures overall socio-environmental neighborhood profile.  {Panel B (spatial visualization and sampled CBGs):} Choropleths of ACS poverty, GEIE canopy, and a composite sustainability index across sampled neighborhoods Phoenix MSA. Patterns show higher poverty and lower canopy in central Phoenix, with greener, lower-poverty conditions in suburban areas. {Panel C (models and predictive performance):} Right schematic shows the two GSV pipelines—GPT-4o via structured prompts (For prompts, see Appendix~\ref{prompts}) and ResNet semantic segmentation that quantifies pixel shares for urban features \cite{zhou2019semantic,zhang2018representing}. Left scatterplots compare GSV--GPT-4o predictions with authoritative estimates (ACS/GEIE) and GSV--ResNet predictions, respectively.  Group-specific OLS fits plotted for {Redlined}, {Ideal}, and {Stable/Declining} CBGs, corresponding to groupings used in the main analyses: {Treatment, Redlined} = HOLC--D;  {Main Comparison, Ideal} = HOLC--A + affluent non-rated Gilbert CBGs; and  {Second Comparison, Stable/Declining} = HOLC B/C + unrated Phoenix CBGs where we expect the estimated treatment effects to be muted compared to the main comparison group.
\end{figure}

\begin{figure}[!htbp]

\begin{center}\includegraphics[width=1\linewidth]{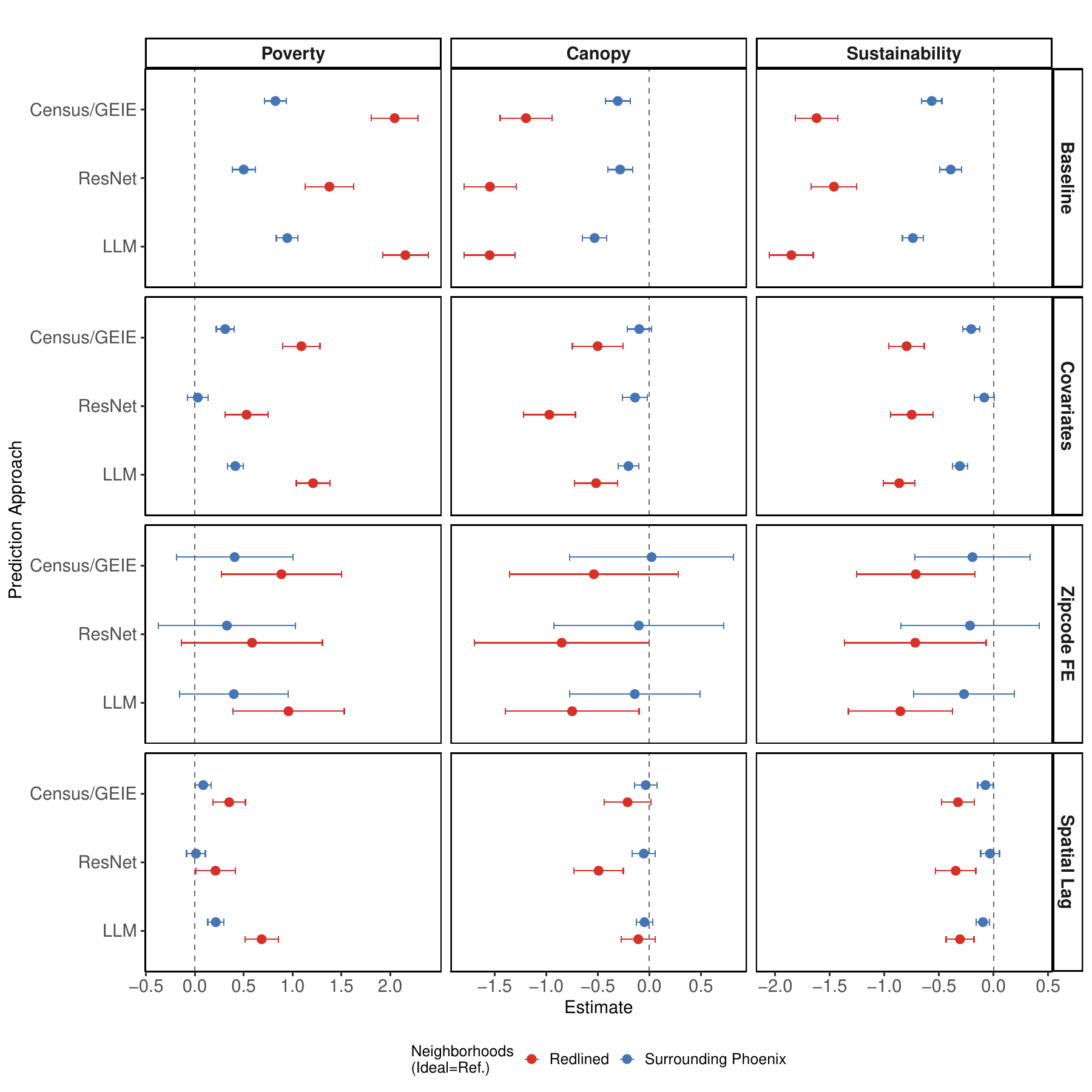} \end{center}

\caption{Estimating Socio-Environmental Legacy Effects of Redlining for Urban Neighborhoods}
\label{fig:FigPolEvalresults}
\scriptsize\textit{Notes:} Table results report results based on variations of the main Eq. \ref{eq:core-m}.  Points and 95\% confidence intervals report standardized coefficients for HOLC--D (redlined) CBGs relative to two benchmarks: {Ideal} (HOLC--A in Phoenix plus affluent, non-rated Gilbert neighborhoods; \textcolor{red}{red}) and {Stable/Declining} (HOLC B/C and surrounding unrated Phoenix neighborhoods; \textcolor{blue}{blue}). Outcomes are Poverty, Tree Canopy, and the composite sustainability index (SI), each estimated under four specifications (Baseline, Covariates, zip code FE, Spatial Lag) and derived from three sources (ACS poverty / GEIE canopy; GSV–GPT-4o; GSV–ResNet). {95\% CIs:} Baseline, Covariates, and zip code FE models use zipcode clustered robust SEs; Spatial Lag (SAR) models report model-based SEs.
\end{figure}

\begin{figure}[!htbp]
\begin{center}\includegraphics[width=1\linewidth]{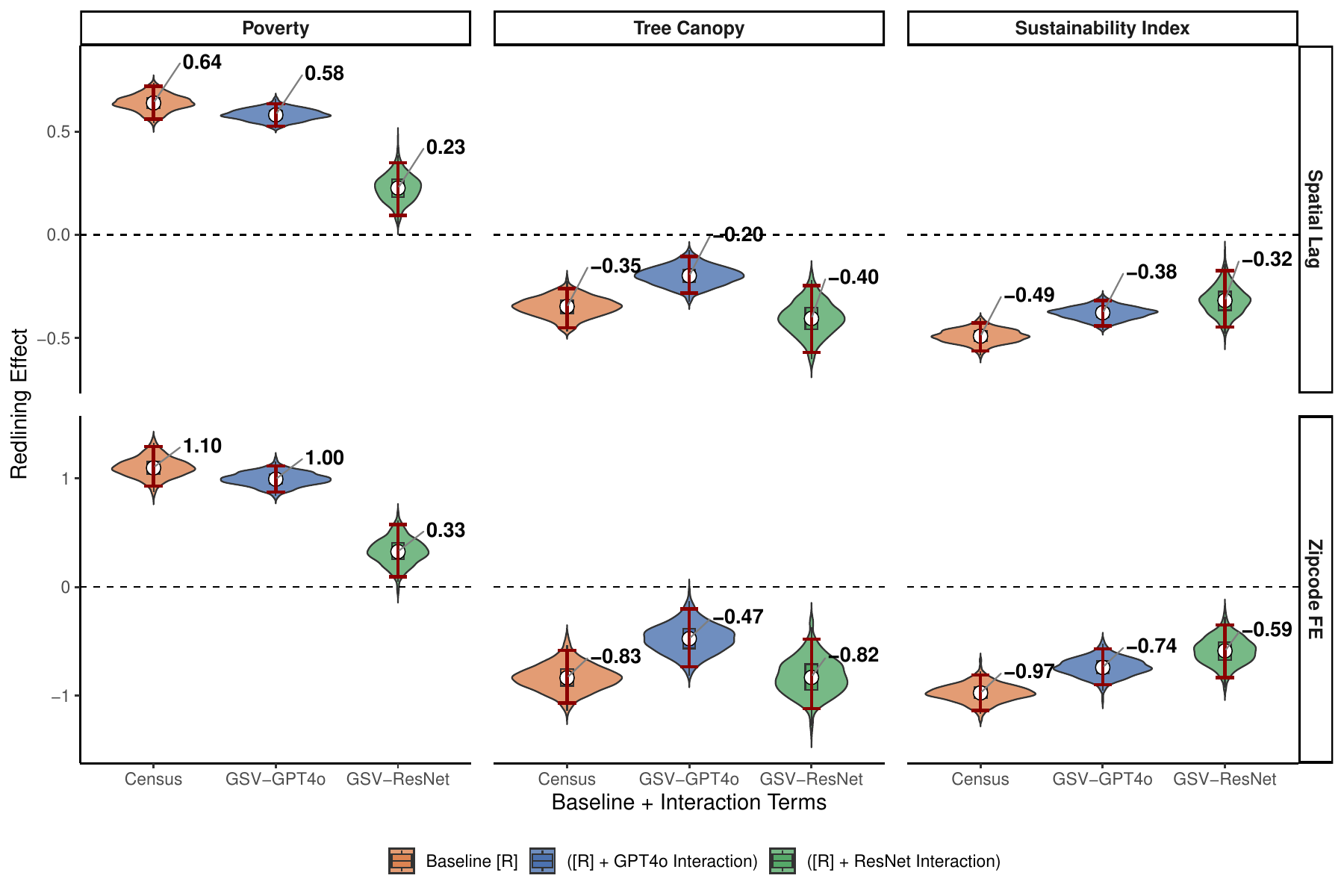} \end{center}
\caption{Bootstrapped Stacked Regression with Redlining x Method Interaction Effects}
\label{fig:Bootstrap_Interaction_Effects}
\vspace{.05in}
\scriptsize\textit{Notes:} The violin plots in Fig.~\ref{fig:Bootstrap_Interaction_Effects} derive from Eq.~\ref{eq:stacked}, displaying empirical bootstrap distributions of \(\delta^{(k)}\) for each approach \(k\) and specification \(\Omega\). Numeric labels above each violin report the bootstrap means of \(\delta^{(k)}\), and red vertical bars mark {percentile} 95\% confidence intervals (2.5th–97.5th percentiles across \(B=500\) draws). The {baseline} redlining coefficient \([R]\) is the estimated effect when outcomes are measured with the authoritative source (ACS for poverty; GEIE for canopy). The total effect for each alternative method equals the baseline plus its interaction with redlining: for GPT-4o, \([R] + [R \times \text{GPT-4o}]\); for ResNet, \([R] + [R \times \text{ResNet}]\). Each violin shows the distribution of coefficient estimates ($\beta$) from 500 CBG-level nonparametric bootstrap resamples.  All outcomes are standardized (z-scored) at the CBG level (coefficients in SD units).

\end{figure}

\begin{figure}[!htbp]

\begin{center}\includegraphics[width=1\linewidth]{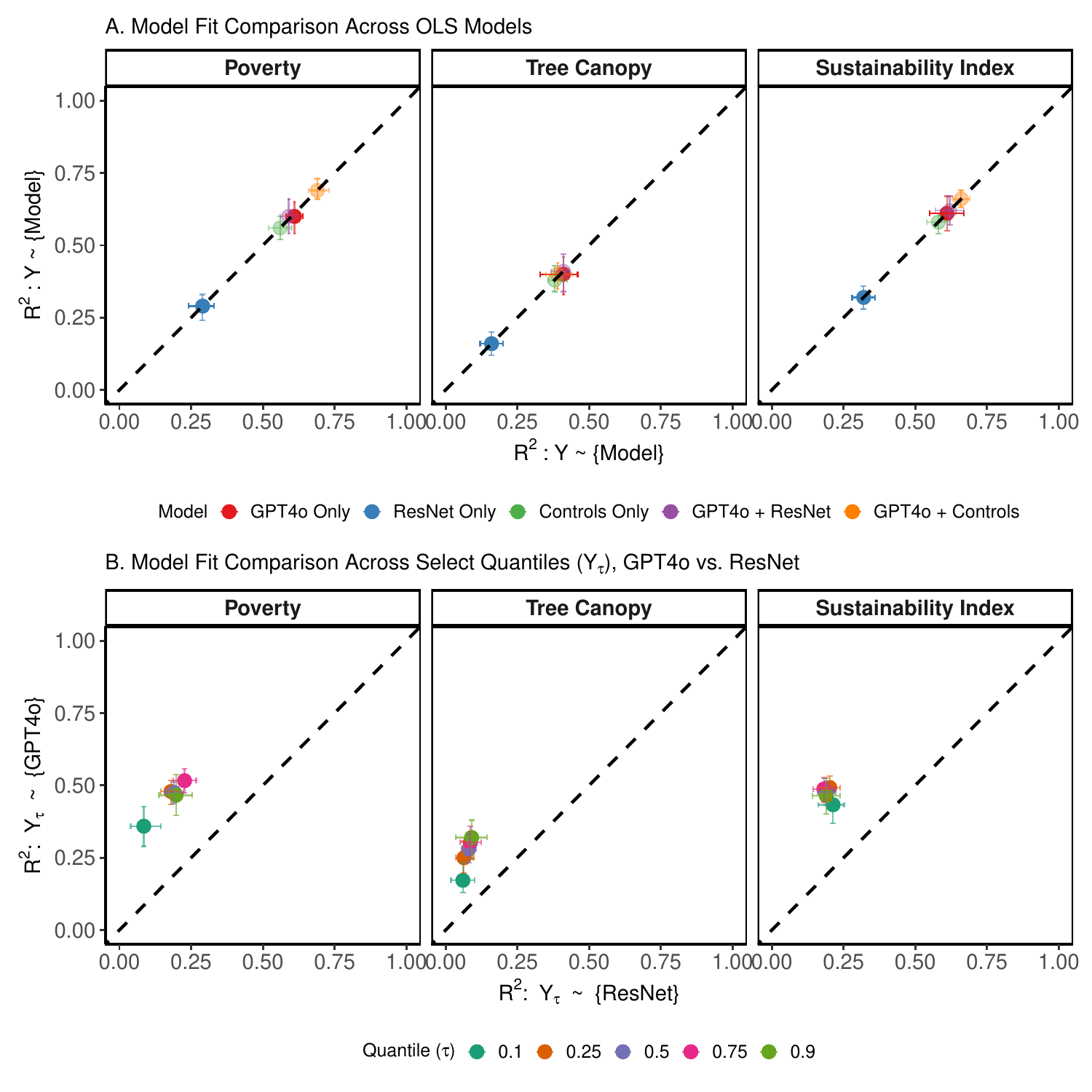} \end{center}

\caption{Model Fit Comparison for GSV--GPT-4o versus GSV-ResNet}
\label{fig:rsquared_analysis}
\vspace{.05in}\scriptsize\textit{Notes:} Each panel summarizes explanatory power for predicting three authoritative outcomes at the CBG level: ACS poverty, GEIE tree canopy, and the composite Sustainability Index (SI $= z(\text{canopy}) - z(\text{poverty})$). {Panel A:} $R^2$ from OLS regressions across five specifications: GPT--4o only, ResNet only, controls only (ACS demographic/socioeconomic covariates), GPT--4o + ResNet, and GPT--4o + controls. Points are bootstrap means (common CBG sample; $B=1000$); error bars show percentile 95\% confidence intervals. {Panel B:} Pseudo-$R^2$ from $\tau$--quantile regressions at selected quantiles ($\tau$), comparing GPT--4o (vertical) to ResNet (horizontal). Each point is the pseudo-$R^2$ for a given $\tau$; the 45$^\circ$ line indicates parity. Pseudo-$R^2$ values are computed from quantile regressions of the authoritative outcome (ACS poverty, GEIE canopy, or SI) on each method's prediction. Bands/intervals reflect percentile 95\% CIs from $B=500$ CBG--level bootstrap resamples.
\end{figure}

We estimate treatment effects of redlining—defined as neighborhoods classified by the Home Owners’ Loan Corporation (HOLC) as ``D'' (hazardous). The HOLC polygons were obtained from the University of Richmond, based on digitized maps as part of the Mapping Inequality Project. Our Phoenix sample contains 27 HOLC--D CBGs. We construct two main reference groups from the official HOLC rating system: (i) an {Ideal} (HOLC--A) group and (ii) a {Stable/Declining} (HOLC--B/C) group. Because the number of rated CBGs in Phoenix is limited, we augment these groups with additional CBGs that were unrated by HOLC but exhibit neighborhood characteristcs that closely resemble their rated counterparts (Fig. \ref{fig:fig1_combinedmap}). 

Specifically, the Ideal group combines the city’s single HOLC--A CBG with 149 affluent, unrated CBGs in nearby Gilbert ({Pseudo--A}). The Stable/Declining group combines the 49 HOLC--B/C CBGs with 919 unrated Phoenix CBGs ({Pseudo--B/C}).  This augmentation  is motivated by the need to increase effective sample sizes in order to improve precision, reducing the risk that method-specific estimates reflect low statistical power. 

Bivariate summaries indicate that redlined (HOLC--D) and Stable/Declining (HOLC--B/C) neighborhoods do not differ statistically on most outcomes or covariates (the notable exception is population density), whereas redlined vs. Ideal (HOLC--A) neighborhoods differ on {every} variable, with redlined areas exhibiting markedly higher poverty and lower canopy across measurement approaches (see Fig.~\ref{fig:fig1_combinedmap}; Table~\ref{tab:summary_by_holc}). Accordingly, we expect smaller redlining effects when comparing D to B/C than to A, $\Delta_{D\text{--}BC} \;<\; \Delta_{D\text{--}A}$, \text{where } 
\begin{equation}
\Delta_{D\text{--}A} \equiv \mathbb{E}[Y \mid \text{HOLC--D}] - \mathbb{E}[Y \mid \text{HOLC--A}],
\Delta_{D\text{--}BC} \equiv \mathbb{E}[Y \mid \text{HOLC--D}] - \mathbb{E}[Y \mid \text{HOLC--B/C}]
\end{equation}

Observing this attenuation serves as an internal robustness (placebo‑style) check that the estimated redlining effects are credible rather than artifacts of measurement, comparison‑group construction, or model specification.

\subsubsection*{Econometric Specifications}
For each measurement approach 
$m \in \{\text{ACS/GEIE},\,\text{GSV--GPT-4o},\,\text{GSV--ResNet}\}$, 
we estimate a variant of the core specification:
\begin{equation}
\label{eq:core-m}
Y_{i}^{(m)} \;=\; \delta^{(m)}\,\text{Redlined}_{i} 
\;+\; X_i' \beta^{(m)} 
\;+\; \mu_{z(i)}^{(m)} 
\;+\; \rho^{(m)}\,(W Y^{(m)})_i 
\;+\; \varepsilon_{i}^{(m)} ,
\end{equation}
where $i$ indexes CBGs, $z(i)$ denotes zip code, 
$X_i$ are covariates from the ACS, $\mu_{z(i)}^{(m)}$ are zip code 
fixed effects, $W$ is a queen–contiguity weights matrix (shared across $m$), 
$\rho^{(m)}$ is the spatial-lag parameter, and $\delta^{(m)}$ is the 
redlining treatment effect under approach $m$. All outcomes $Y^{(m)}$ are 
standardized. Corresponding to the presentation of results 
(Fig.~\ref{fig:FigPolEvalresults}; 
Tables~\ref{HOLC coefficient estimates – Poverty}–\ref{HOLC coefficient estimates – Sustainability}), 
we estimate treatment effects from progressively saturated model variants: 
(i) unadjusted differences ($X_i=\mathbf{0}$, $\mu_{z(i)}=0$, $\rho=0$); 
(ii) covariate-adjusted ($X_i \neq \mathbf{0}$, $\mu_{z(i)}=0$, $\rho=0$); 
(iii) zip code FE ($X_i \neq \mathbf{0}$, $\mu_{z(i)}\neq 0$, $\rho=0$); and 
(iv) SAR ($X_i \neq \mathbf{0}$, $\mu_{z(i)}=0$, $\rho\neq 0$).

\subsubsection*{Stacked Regression Approach}

To compare measurement approaches \(m \in \{\text{ACS/GEIE},\,\text{GSV--GPT-4o},\,\text{GSV--ResNet}\}\) more formally through equivalence testing, we pool all three outcome series for the same CBGs and estimate the {zip code} FE and SAR variants defined above (denoted by \(\Omega\)):
\begin{equation}
\label{eq:stacked}
Y_{i,k}\mid\Omega
\;=\;
\delta_{0}\,\text{Redlined}_{i}
\;+\;
\sum_{\substack{m\in\{\text{GSV--GPT-4o},\\ \text{GSV--ResNet}\}}}
\eta_{m}\,\mathbb{I}[k=m]
\;+\;
\sum_{\substack{m\in\{\text{GSV--GPT-4o},\\ \text{GSV--ResNet}\}}}
\theta_{m}\,\big(\text{Redlined}_{i}\times \mathbb{I}[k=m]\big)
\end{equation}

where $i$ indexes CBGs, $k$ indexes approaches, and $\mathbb{I}[k=m]$ is the indicator for approach $m$. The ACS/GEIE approach is the omitted category. Here, $\delta_{0}$ is the baseline redlining effect under the authoritative approach; $\eta_m$ captures systematic level shifts in outcomes associated with approach $m$; and $\theta_m$ captures differences in the redlining effect for approach $m$ relative to ACS/GEIE.  

Uncertainty is quantified with a nonparametric CBG–level cluster bootstrap ($B=500$) applied separately under each specification $\Omega\in\{\text{zip code FE},\text{SAR}\}$. In each bootstrap draw $b$: (1) CBGs are sampled with replacement, retaining all stacked rows for each selected CBG across approaches; (2) Eq.~\eqref{eq:stacked} is re-estimated under $\Omega$, yielding $\delta_{0}^{(b)}$, $\theta_{\text{GSV--GPT-4o}}^{(b)}$, and $\theta_{\text{GSV--ResNet}}^{(b)}$; (3) implied approach-specific effects are formed by addition:
\[
\delta^{\text{ACS/GEIE},(b)}=\delta_{0}^{(b)},\quad
\delta^{\text{GSV--GPT-4o},(b)}=\delta_{0}^{(b)}+\theta_{\text{GSV--GPT-4o}}^{(b)},\quad
\delta^{\text{GSV--ResNet},(b)}=\delta_{0}^{(b)}+\theta_{\text{GSV--ResNet}}^{(b)}.
\]

The violin plots in Fig.~\ref{fig:Bootstrap_Interaction_Effects} show the empirical bootstrap distributions of $\delta^{(k)}$ for each approach $k$, providing nonparametric uncertainty bands that accommodate repeated CBGs in the stacked dataset. Table~\ref{tab:All_interactions} reports results from one stacked-regression design for illustration. The baseline ACS/GEIE redlining effect is given in the first row; interaction terms show deviations under GPT-4o and ResNet. The implied total effects equal the baseline plus their respective interaction, i.e., $\delta^{(m)} \;=\; \delta_{0} \;+\; \theta_{m}$, which are the quantities illustrated in the violin plots.  

Equivalence tests are conducted as $H_{0}:\theta_{m}=0$. We find that $H_{0}:\theta_{\text{GPT-4o}}=0$ cannot be rejected across all outcomes and specifications, implying GPT-4o replicates the ACS/GEIE benchmark. For ResNet, $H_{0}:\theta_{\text{ResNet}}=0$ cannot be rejected for canopy and SI, but is rejected for poverty under both the FE and SAR specifications, consistent with ResNet’s tendency to attenuate poverty effects.


\clearpage
\newpage

\bibliographystyle{unsrt} 

\bibliography{mybibfile}

\clearpage
\newpage

\section{Figures}\label{figures}
\setcounter{page}{1}

\newpage
\clearpage

\setcounter{page}{1}

\section{Supplementary Information}\label{supplementary-information-si}

\newpage
\clearpage
\setcounter{page}{1}

\setcounter{table}{0} \renewcommand{\thetable}{S\arabic{table}} \setcounter{figure}{0} \renewcommand{\thefigure}{S\arabic{figure}}

\renewcommand{\thesubsection}{\Alph{subsection}}
\setcounter{subsection}{0}

\clearpage
\newpage

\subsection{Prompt Design}\label{prompts}

\paragraph{Global preamble}
You are an urban auditor producing an {area-level} proxy from a single street-level image. Use {only} visible, non-sensitive {SVFs} (eg., structures, lots, roads, vegetation, street furniture, lighting). {Do not} infer characteristics of identifiable people. All numeric outputs must follow the stated ranges and be {rounded to two decimals}. Return {only valid JSON} matching the prompt's schema.

\paragraph{Common output rule.}
Each prompt must return {only} the JSON object described in its "Answer (JSON)" block (no extra text).

\newenvironment{promptframe}{%
  \begin{framed}\setlength{\parindent}{0pt}\setlength{\parskip}{6pt}%
}{\end{framed}}

\begin{promptframe}
    
\subsubsection*{Prompt 1: Housing Structure and Facade Indicators}
\begin{quote}
\textbf{Task.} Identify the primary residential structure type (if any) and facade-level maintenance indicators (non-sensitive, visual cues only).

\textbf{Allowed structure types (choose one).}
\begin{lstlisting}
"single_family_detached","duplex","mobile_home","apartment",
"townhouse","mixed_use","other","unknown","none_visible"
\end{lstlisting}

\textbf{Answer.}
\begin{verbatim}
{
  "structure_type": "<one allowed value>",
  "facade_indicators": ["<zero or more canonical tokens>"],
  "n_facade_indicators": <integer >= 0>,
  "notes": "<<=100 chars, optional>"
}
\end{verbatim}
\end{quote}
\end{promptframe}

\subsubsection*{Prompt 2: Environmental Deprivation Indicators}

\begin{promptframe}

\begin{quote}
\textbf{Task.} Record neighborhood-scale environmental indicators associated with lower infrastructure quality or maintenance (non-sensitive, visual cues only). Also record whether overhead canopy elements are present.

{Canonical environmental indicators (choose zero or more).}
\begin{lstlisting}
"dirt_lot","overgrowth","debris","landscaping_absent",
"vehicle_damaged","vehicle_abandoned","very_old_vehicle",
"driveway_broken","cracked_sidewalk","poor_lighting",
"potholes","window_bars","perimeter_fence","clutter_disrepair"
\end{lstlisting}

{Canopy indicator tokens (choose zero or more).}
\begin{lstlisting}
"tree","palm","large_shrub"
\end{lstlisting}

\textbf{Answer.}
\begin{lstlisting}
{
  "env_indicators": ["<zero or more canonical tokens from the list above>"],
  "n_env_indicators": <integer >= 0>,
  "canopy_indicators": ["<zero or more of: tree, palm, large_shrub>"],
  "n_canopy_indicators": <integer >= 0>,
  "notes": "<<=100 chars, optional>"
}
\end{lstlisting}
\end{quote}
\end{promptframe}

\subsubsection*{Prompt 3: Tree Canopy Coverage (Chained Estimation)}

\begin{promptframe}

\begin{quote}
\textbf{Task.} Using the provided indicators, estimate the {local-scene} share of visible area covered by {overhead canopy only} (mature trees, palms, large shrubs). Exclude grass, small bushes, flowerbeds, and groundcover.

{You will be provided with (as variables):}
\begin{lstlisting}
"canopy_indicators": ["tree" | "palm" | "large_shrub", ...],
"n_canopy_indicators": <int>
\end{lstlisting}

{Calibration anchors (choose one band, then give a numeric).}
\begin{lstlisting}
very_low (0.00-0.20), low (0.20-0.40), moderate (0.40-0.60),
high (0.60-0.80), very_high (0.80-1.00), unknown
\end{lstlisting}

{Critical numeric rule.}
The numeric estimate \verb|canopy_share_0_1| must be a continuous value {inside} the chosen band (e.g., if \verb|low|, use 0.21-0.39). {Do not} return band cutpoints (0.20, 0.40, etc.) as placeholders. Use exact \verb|0.00| {only if} the scene unambiguously shows no canopy (\verb|"n_canopy_indicators": 0|).

\textbf{Answer.}
\begin{lstlisting}
{
  "canopy_band": "very_low" | "low" | "moderate" | "high" | "very_high" | "unknown",
  "canopy_share_0_1": <float in [0.00, 1.00] or null>,
  "notes": "<<=100 chars, optional>"
}
\end{lstlisting}

{Rounding.} Report \verb|canopy_share_0_1| to two decimals. If information is insufficient, set \verb|"canopy_band":"unknown"| and \verb|"canopy_share_0_1": null|.
\end{quote}
\end{promptframe}

\subsubsection*{Prompt 4: Area-Level Poverty Prevalence Proxy (Chained Re-Estimation)}

\begin{promptframe}

\begin{quote}
\textbf{Task.} Using the provided structure type and the facade/environmental indicators, produce a {local-scene, area-level} proxy for the {share of households below U.S. federal poverty thresholds (2023)}. Use {only} visible built-environment cues; {do not} infer person-level attributes.

{You will be provided with (as variables):}
\begin{lstlisting}
"structure_type": "<one of the allowed types>",
"facade_indicators": ["<tokens>"], "n_facade_indicators": <int>,
"env_indicators": ["<tokens>"], "n_env_indicators": <int>
\end{lstlisting}

{Calibration anchors (choose one band, then give a numeric).}
\begin{lstlisting}
very_low (0.00-0.20), low (0.20-0.40), moderate (0.40-0.60),
high (0.60-0.80), very_high (0.80-1.00), unknown
\end{lstlisting}

{Critical numeric rule.}
The numeric estimate \verb|poverty_proxy_0_1| must be a continuous value {inside} the chosen band (e.g., if \verb|moderate|, use 0.41-0.59). {Do not} return band midpoints or cutpoints as placeholders. Use exact \verb|0.00| {only if} no facade or environmental indicators of deprivation are present (\verb|"n_facade_indicators": 0, "n_env_indicators": 0|).

\textbf{Answer.}
\begin{lstlisting}
{
  "poverty_band": "very_low" | "low" | "moderate" | "high" | "very_high" | "unknown",
  "poverty_proxy_0_1": <float in [0.00, 1.00] or null>,
  "evidence_counts": {
    "n_facade_indicators": <int>,
    "n_env_indicators": <int>
  },
  "notes": "<<=100 chars, optional>"
}
\end{lstlisting}

{Rounding.} Report \verb|poverty_proxy_0_1| to two decimals. If information is insufficient, set \verb|"poverty_band":"unknown"| and \verb|"poverty_proxy_0_1": null|.
\end{quote}
\end{promptframe}

\newpage

\setcounter{page}{1}

\subsection*{SI Figures}\label{appnFig}
\addcontentsline{toc}{subsection}{SI Figures}

\begin{figure}[!htbp]

\begin{center}\includegraphics[width=1\linewidth]{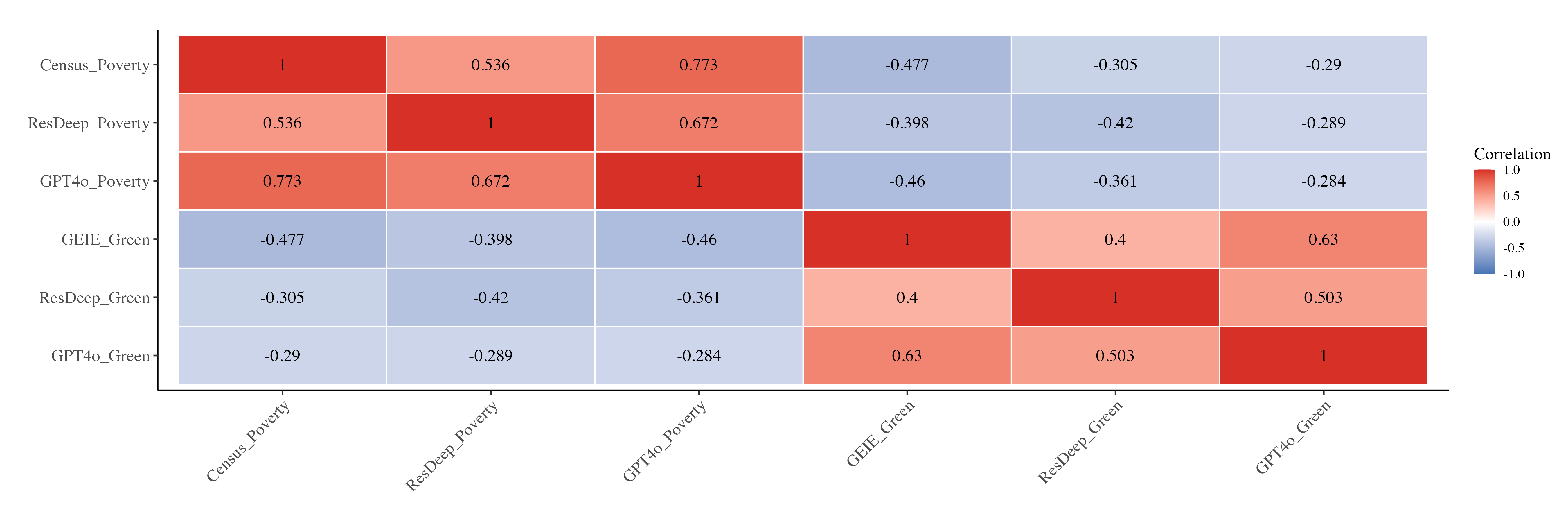} \end{center}

\caption{Correlation Heatmap: Census, ResNet, and GPT4o Predictions.}
\label{fig:correlation_heatmap_ordered}
\vspace{.05in}\scriptsize\textit{Notes: Heatmap illustrating pairwise correlations among poverty and tree canopy coverage metrics. Official poverty estimates obtained from ACS, U.S. Census Bureau; canopy data from GEIE. ResNet predictions derived from pixel-level semantic segmentation trained on MIT ADE20K dataset; GPT4o predictions from visual inference prompt applied to GSV imagery. Positive correlations (red) signify alignment between metrics, whereas negative correlations (blue) indicate inverse relationships, particularly between poverty and canopy coverage. Census Size-weighted predictions aggregated at CBG level across Maricopa County, Arizona.}
\end{figure}

\newpage
\setcounter{page}{1}

\subsection*{SI Tables}\label{appn}
\addcontentsline{toc}{subsection}{SI Tables}

\begin{table}[!htbp]
\centering
\caption{\label{tab:summary_by_holc}Summary Statistics by HOLC Grade}
\footnotesize
\begin{threeparttable}
\begin{tabular}{l @{\extracolsep{\fill}} ccc}
\toprule
 & Ideal (HOLC--A) & Stable/Declining (HOLC--B/C) & Hazardous (HOLC--D, Redlined)\\
\midrule
\multicolumn{4}{l}{\textbf{Sample}}\\
\hspace{1em}GSV Panoramas (\#) & 3750 & 21810 & 725\\
\hspace{1em}CBG (\#) & 150 & 980 & 29\\
\addlinespace[0.3em]
\multicolumn{4}{l}{\textbf{Outcomes}}\\
\hspace{1em}Poverty (Census) & 0.131 (0.096) & 0.338 (0.217) & 0.616* (0.226)\\
\hspace{1em}Tree Canopy (GEIE) & 0.147 (0.042) & 0.121 (0.043) & 0.074* (0.031)\\
\hspace{1em}GPT4o Poverty & 0.042 (0.025) & 0.107 (0.085) & 0.201* (0.04)\\
\hspace{1em}GPT4o Canopy & 0.043 (0.01) & 0.038 (0.016) & 0.025* (0.009)\\
\hspace{1em}ResNet Poverty & 0.266 (0.091) & 0.324 (0.111) & 0.424* (0.081)\\
\hspace{1em}ResNet Canopy & 0.129 (0.017) & 0.123 (0.018) & 0.1* (0.018)\\
\addlinespace[0.3em]
\multicolumn{4}{l}{\textbf{Controls}}\\
\hspace{1em}Population Density & 2.125 (0.068) & 2.151 (0.122) & 2.075** (0.138)\\
\hspace{1em}Dependency Rate & 0.33 (0.12) & 0.293 (0.13) & 0.283 (0.187)\\
\hspace{1em}Linguistic Isolation & 0.026 (0.028) & 0.132 (0.156) & 0.152* (0.116)\\
\hspace{1em}Black (\%) & 0.033 (0.049) & 0.07 (0.097) & 0.117* (0.126)\\
\hspace{1em}Hispanic (\%) & 0.171 (0.102) & 0.389 (0.286) & 0.582* (0.246)\\
\hspace{1em}Asian (\%) & 0.065 (0.069) & 0.039 (0.065) & 0.013* (0.025)\\
\hspace{1em}College Educated (\%) & 0.196 (0.058) & 0.144 (0.105) & 0.062* (0.048)\\

\bottomrule
\end{tabular}
\begin{tablenotes}
\item \scriptsize\textit{Notes:}: Means with standard deviations in parentheses. The Ideal group includes HOLC--A graded neighborhoods and non-graded Gilbert neighborhoods. The Fair group includes all non-redlined neighborhoods in Phoenix. * indicates a statistically significant difference in means between D and A. ** indicates a significant difference between D and both A and B/C ($p < 0.05$). 
\end{tablenotes}
\end{threeparttable}
\end{table}

\begin{table}[!htbp]
\centering
\caption{Redlining Effects on Poverty, by Measurement Appraoch}

\label{HOLC coefficient estimates – Poverty}
\begin{threeparttable}
\footnotesize
\begin{tabular}{lcccc}
\toprule
 & Baseline & Covariates & ZIP-code FE & Spatial Lag \\
\midrule

{\bfseries{A. Ref=Ideal CBGs}}\\

Redlined (Census/GEIE) & $2.046^{*}$ & $1.091^{*}$ & $0.886^{*}$ & $0.352^{*}$ \\
                       & $(0.122)$   & $(0.097)$   & $(0.313)$   & $(0.085)$   \\
Redlined (GPT-4o)      & $2.156^{*}$ & $1.211^{*}$ & $0.959^{*}$ & $0.684^{*}$ \\
                       & $(0.119)$   & $(0.087)$   & $(0.291)$   & $(0.087)$   \\
Redlined (ResNet)      & $1.378^{*}$ & $0.530^{*}$ & $0.585$ & $0.212^{*}$ \\
                       & $(0.127)$   & $(0.112)$   & $(0.368)$   & $(0.103)$   \\

{\bfseries{B. Ref=Stable/Declining CBGs}}\\

Redlined (Census/GEIE) & $0.825^{*}$ & $0.310^{*}$ & $0.408$ & $0.086^{*}$ \\
                       & $(0.058)$   & $(0.047)$   & $(0.305)$   & $(0.040)$   \\
Redlined (GPT-4o)      & $0.946^{*}$ & $0.414^{*}$ & $0.400$ & $0.214^{*}$ \\
                       & $(0.057)$   & $(0.042)$   & $(0.283)$   & $(0.042)$   \\
Redlined (ResNet)      & $0.501^{*}$ & $0.030$ & $0.328$ & $0.011$ \\
                       & $(0.060)$   & $(0.054)$   & $(0.358)$   & $(0.049)$   \\
\bottomrule
\end{tabular}
\begin{tablenotes}[flushleft]
\item \scriptsize\textit{Notes:} $^{*}$ denotes statistical significance at the 0.05 level. Entries are standardized treatment-effect estimates (SE in parentheses) for HOLC-D (“Redlined”) relative to two reference groups: \emph{Ideal} (HOLC-A plus affluent, non-HOLC-rated suburban CBGs) and \emph{Stable/Declining} (HOLC B/C plus unrated Phoenix CBGs). Poverty outcomes are constructed from three sources: authoritative (ACS), MLLM-based estimates from GSV (GPT-4o), and segmentation-based estimates from GSV (ResNet). Standard errors are robust and clustered at the zipcode level.
\end{tablenotes}
\end{threeparttable}
\end{table}

\begin{table}[!htbp]
\centering
\caption{Redlining Effects on Tree Canopy, by Measurement Appraoch}

\label{HOLC coefficient estimates – Canopy}
\begin{threeparttable}
\footnotesize
\begin{tabular}{lcccc}
\toprule
 & Baseline & Covariates & ZIP-code FE & Spatial Lag \\
\midrule

{\bfseries{A. Ref=Ideal CBGs}}\\

Redlined (GEIE)            & $-1.196^{*}$  & $-0.501^{*}$  & $-0.538$  & $-0.212^{*}$  \\
                                  & $(0.128)$     & $(0.125)$     & $(0.417)$     & $(0.116)$     \\
Redlined (GPT4o)                  & $-1.549^{*}$  & $-0.517^{*}$  & $-0.748^{*}$  & $-0.107^{*}$  \\
                                  & $(0.126)$     & $(0.107)$     & $(0.331)$     & $(0.085)$     \\
Redlined (ResNet)                 & $-1.545^{*}$  & $-0.970^{*}$  & $-0.849$  & $-0.492^{*}$  \\
                                  & $(0.129)$     & $(0.129)$     & $(0.432)$     & $(0.123)$     \\

{\bfseries{B. Ref=Stable/Declining CBGs}}\\

Redlined (GEIE)            & $-0.306^{*}$  & $-0.098$  & $0.022$   & $-0.035$  \\
                                  & $(0.061)$     & $(0.060)$     & $(0.405)$     & $(0.055)$     \\
Redlined (GPT4o)                  & $-0.531^{*}$  & $-0.203^{*}$  & $-0.141$  & $-0.047$  \\
                                  & $(0.060)$     & $(0.051)$     & $(0.322)$     & $(0.040)$     \\
Redlined (ResNet)                 & $-0.283^{*}$  & $-0.139^{*}$  & $-0.102$  & $-0.054$  \\
                                  & $(0.062)$     & $(0.062)$     & $(0.420)$     & $(0.057)$     \\
\bottomrule
\end{tabular}
\begin{tablenotes}[flushleft]
\item \scriptsize\textit{Notes:} $^{*}$ denotes statistical significance at the 0.05 level. Entries are standardized treatment-effect estimates (SE in parentheses) for HOLC-D (“Redlined”) relative to two reference groups: \emph{Ideal} (HOLC-A plus affluent, non-HOLC-rated suburban CBGs) and \emph{Stable/Declining} (HOLC B/C plus unrated Phoenix CBGs). Tree-canopy outcomes are constructed from three sources: authoritative (GEIE), MLLM-based estimates from GSV (GPT-4o), and segmentation-based estimates from GSV (ResNet). Standard errors are robust and clustered at the zipcode level.
\end{tablenotes}
\end{threeparttable}
\end{table}

\begin{table}[!htbp]
\centering
\caption{Redlining Effects on Sustainability Index, by Measurement Appraoch}

\label{HOLC coefficient estimates – Sustainability}
\begin{threeparttable}
\footnotesize
\begin{tabular}{lcccc}
\toprule
 & Baseline & Covariates & ZIP-code FE & Spatial Lag \\
\midrule

{\bfseries{A. Ref=Ideal CBGs}}\\
Redlined (Census/GEIE)            & $-1.621^{*}$ & $-0.796^{*}$ & $-0.712^{*}$ & $-0.326^{*}$ \\
                                  & $(0.099)$    & $(0.083)$    & $(0.277)$    & $(0.077)$    \\
Redlined (GPT-4o)                 & $-1.852^{*}$ & $-0.864^{*}$ & $-0.854^{*}$ & $-0.307^{*}$ \\
                                  & $(0.103)$    & $(0.074)$    & $(0.243)$    & $(0.065)$    \\
Redlined (ResNet)                 & $-1.462^{*}$ & $-0.750^{*}$ & $-0.717^{*}$ & $-0.347^{*}$ \\
                                  & $(0.106)$    & $(0.099)$    & $(0.332)$    & $(0.094)$    \\
{\bfseries{B. Ref=Stable/Declining CBGs}}\\

Surrounding Phoenix (Census/GEIE) & $-0.566^{*}$ & $-0.204^{*}$ & $-0.193$ & $-0.074^{*}$ \\
                                  & $(0.047)$    & $(0.040)$    & $(0.269)$    & $(0.036)$    \\
Surrounding Phoenix (GPT-4o)      & $-0.739^{*}$ & $-0.309^{*}$ & $-0.271$ & $-0.097^{*}$ \\
                                  & $(0.049)$    & $(0.035)$    & $(0.236)$    & $(0.031)$    \\
Surrounding Phoenix (ResNet)      & $-0.392^{*}$ & $-0.085$ & $-0.215$ & $-0.031$ \\
                                  & $(0.051)$    & $(0.048)$    & $(0.322)$    & $(0.044)$    \\
\bottomrule
\end{tabular}
\begin{tablenotes}[flushleft]
\item \scriptsize\textit{Notes:} $^{*}$ denotes statistical significance at the 0.05 level. Entries are standardized treatment-effect estimates (SE in parentheses) for HOLC-D (“Redlined”) relative to two reference groups: \emph{Ideal} (HOLC-A plus affluent, non-HOLC-rated suburban CBGs) and \emph{Stable/Declining} (HOLC B/C plus unrated Phoenix CBGs). Sustainability index outcomes are constructed by taking the standardized difference between poverty and tree canopy from three sources: authoritative (ACS/GEIE), MLLM-based estimates from GSV (GPT-4o), and segmentation-based estimates from GSV (ResNet). Standard errors are robust and clustered at the zipcode level.
\end{tablenotes}
\end{threeparttable}
\end{table}

\begin{table}[htbp]
\caption{Regression Models Predicting ACS-based Poverty Rates}
\label{tab:poverty_regression}
\centering
\footnotesize
  \begin{threeparttable}
    \begin{tabular*}{\textwidth}{l @{\extracolsep{\fill}} c c c c c} 
\toprule
 & GPT4o  & ResNet  & Both GPT4o  & Demographic  & All \\
 &  Only &  Only & + ResNet &  Only &  \\

\midrule
Constant                & $0.102^{*}$ & $-0.021$    & $0.089^{*}$ & $-0.054$     & $-0.108^{*}$ \\
                        & $(0.007)$   & $(0.017)$   & $(0.013)$   & $(0.055)$    & $(0.046)$    \\
GPT4o Poverty           & $2.199^{*}$ &             & $2.140^{*}$ &              & $1.449^{*}$  \\
                        & $(0.053)$   &             & $(0.072)$   &              & $(0.064)$    \\
ResNet Poverty          &             & $1.075^{*}$ & $0.062$     &              &              \\
                        &             & $(0.050)$   & $(0.051)$   &              &              \\
Population Density (ln) &             &             &             & $0.034^{*}$  & $0.025^{*}$  \\
                        &             &             &             & $(0.007)$    & $(0.006)$    \\
Dependency Rate         &             &             &             & $0.105^{*}$  & $0.142^{*}$  \\
                        &             &             &             & $(0.034)$    & $(0.029)$    \\
Linguistic Isolation    &             &             &             & $0.176^{*}$  & $0.085^{*}$  \\
                        &             &             &             & $(0.035)$    & $(0.030)$    \\
Black (\%)              &             &             &             & $0.437^{*}$  & $0.307^{*}$  \\
                        &             &             &             & $(0.047)$    & $(0.040)$    \\
Hispanic (\%)           &             &             &             & $0.306^{*}$  & $0.151^{*}$  \\
                        &             &             &             & $(0.025)$    & $(0.022)$    \\
Asian (\%)              &             &             &             & $-0.024$     & $0.011$      \\
                        &             &             &             & $(0.070)$    & $(0.058)$    \\
College Educated (\%)   &             &             &             & $-0.492^{*}$ & $-0.257^{*}$ \\
                        &             &             &             & $(0.062)$    & $(0.053)$    \\
\midrule
Num. obs.               & $1145$      & $1145$      & $1145$      & $1145$       & $1145$       \\
Adj. R$^2$ & $0.597$     & $0.287$     & $0.597$     & $0.553$      & $0.690$      \\
\bottomrule
\end{tabular*}
\begin{tablenotes}[flushleft]
\item \scriptsize\textit{Notes:} $^{*}p<0.05$. Dependent variable is the ACS 2023 poverty rate at the census block-group (CBG) level (Phoenix MSA, $N=1{,}145$).  GPT-4o Poverty and ResNet Poverty are outcome estimates inferred from GSV imagery using a multimodal LLM and a pixel-based semantic-segmentation pipeline, respectively.  Columns compare explanatory power across specifications: GPT4o Only uses the GPT-4o measure alone; ResNet Only uses the ResNet measure alone; Both GPT4o + ResNet includes both image-derived measures; Demographic Only includes the listed covariates; All includes GPT-4o plus demographics.  Estimation is by OLS with robust SEs in parentheses. 
\end{tablenotes}

  \end{threeparttable}
\end{table}

\begin{table}[htbp]
\caption{Regression Models Predicting GEIE-based Tree Canopy Coverage}
\label{tab:tree_regression}
\centering
\footnotesize
  \begin{threeparttable}
    \begin{tabular*}{\textwidth}{l @{\extracolsep{\fill}} c c c c c} 
\toprule
 & GPT4o  & ResNet  & Both GPT4o  & Demographic  & All \\
 &  Only &  Only & + ResNet &  Only &  \\
\midrule
Constant                & $0.050^{*}$ & $0.004$     & $0.023^{*}$ & $0.001$      & $-0.004$     \\
                        & $(0.003)$   & $(0.008)$   & $(0.007)$   & $(0.013)$    & $(0.011)$    \\
GPT4o Canopy            & $1.923^{*}$ &             & $1.751^{*}$ &              & $1.415^{*}$  \\
                        & $(0.070)$   &             & $(0.081)$   &              & $(0.065)$    \\
ResNet Canopy           &             & $0.969^{*}$ & $0.271^{*}$ &              &              \\
                        &             & $(0.066)$   & $(0.064)$   &              &              \\
Population Density (ln) &             &             &             & $0.015^{*}$  & $0.009^{*}$  \\
                        &             &             &             & $(0.002)$    & $(0.001)$    \\
Dependency Rate         &             &             &             & $0.030^{*}$  & $0.016^{*}$  \\
                        &             &             &             & $(0.008)$    & $(0.007)$    \\
Linguistic Isolation    &             &             &             & $-0.020^{*}$ & $-0.008$     \\
                        &             &             &             & $(0.008)$    & $(0.007)$    \\
Black (\%)              &             &             &             & $-0.071^{*}$ & $-0.044^{*}$ \\
                        &             &             &             & $(0.011)$    & $(0.009)$    \\
Hispanic (\%)           &             &             &             & $-0.042^{*}$ & $-0.032^{*}$ \\
                        &             &             &             & $(0.006)$    & $(0.005)$    \\
Asian (\%)              &             &             &             & $-0.044^{*}$ & $-0.027$     \\
                        &             &             &             & $(0.016)$    & $(0.014)$    \\
College Educated (\%)   &             &             &             & $0.150^{*}$  & $0.110^{*}$  \\
                        &             &             &             & $(0.014)$    & $(0.012)$    \\
\midrule
Adj. R$^2$              & $0.396$     & $0.159$     & $0.405$     & $0.377$      & $0.559$      \\
Num. obs.               & $1145$      & $1145$      & $1145$      & $1145$       & $1145$       \\
\bottomrule
\end{tabular*}
\begin{tablenotes}[flushleft]
\item \scriptsize\textit{Notes:} $^{*}p<0.05$. Dependent variable is the GEIE 2023 tree canopy coverage at the census block-group (CBG) level (Phoenix MSA, $N=1{,}145$).  GPT-4o Poverty and ResNet Poverty are outcome estimates inferred from GSV imagery using a multimodal LLM and a pixel-based semantic-segmentation pipeline, respectively.  Columns compare explanatory power across specifications: GPT4o Only uses the GPT-4o measure alone; ResNet Only uses the ResNet measure alone; Both GPT4o + ResNet includes both image-derived measures; Demographic Only includes the listed covariates; All includes GPT-4o plus demographics.  Estimation is by OLS with robust SEs in parentheses. 
\end{tablenotes}

  \end{threeparttable}
\end{table}

\begin{table}[htbp]
\caption{Regression Models Predicting Composite Sustainability Index (SI)}
\label{tab:sustainability_regression}
\centering
\footnotesize
  \begin{threeparttable}
    \begin{tabular*}{\textwidth}{l @{\extracolsep{\fill}} c c c c c} 
\toprule
 & GPT4o  & ResNet  & Both GPT4o  & Demographic  & All \\
 &  Only &  Only & + ResNet &  Only &  \\
\midrule
Constant                & $0.000$     & $0.000$     & $0.000$     & $-0.566^{*}$ & $-0.114$     \\
                        & $(0.016)$   & $(0.021)$   & $(0.016)$   & $(0.213)$    & $(0.172)$    \\
GPT4o SI    & $0.839^{*}$ &             & $0.788^{*}$ &              & $0.543^{*}$  \\
                        & $(0.020)$   &             & $(0.027)$   &              & $(0.022)$    \\
ResNet SI   &             & $0.577^{*}$ & $0.071^{*}$ &              &              \\
                        &             & $(0.025)$   & $(0.025)$   &              &              \\
Population Density (ln) &             &             &             & $0.099^{*}$  & $0.035$      \\
                        &             &             &             & $(0.025)$    & $(0.020)$    \\
Dependency Rate         &             &             &             & $0.108$      & $-0.173$     \\
                        &             &             &             & $(0.131)$    & $(0.106)$    \\
Linguistic Isolation    &             &             &             & $-0.639^{*}$ & $-0.243^{*}$ \\
                        &             &             &             & $(0.136)$    & $(0.110)$    \\
Black (\%)              &             &             &             & $-1.822^{*}$ & $-1.140^{*}$ \\
                        &             &             &             & $(0.182)$    & $(0.148)$    \\
Hispanic (\%)           &             &             &             & $-1.187^{*}$ & $-0.677^{*}$ \\
                        &             &             &             & $(0.094)$    & $(0.078)$    \\
Asian (\%)              &             &             &             & $-0.457$     & $-0.309$     \\
                        &             &             &             & $(0.269)$    & $(0.216)$    \\
College Educated (\%)   &             &             &             & $2.870^{*}$  & $1.746^{*}$  \\
                        &             &             &             & $(0.239)$    & $(0.197)$    \\
\midrule
Adj. R$^2$              & $0.611$     & $0.320$     & $0.613$     & $0.575$      & $0.727$      \\
Num. obs.               & $1145$      & $1145$      & $1145$      & $1145$       & $1145$       \\
\bottomrule
\end{tabular*}
\begin{tablenotes}[flushleft]
\item \scriptsize\textit{Notes:} $^{*}p<0.05$. Dependent variable is the composite Sustainability Index that is the standardized difference between  2023 ACS poverty and GEIE Tree canopy coverage  at the census block-group (CBG) level (Phoenix MSA, $N=1{,}145$).  GPT-4o Poverty and ResNet Poverty are outcome estimates inferred from GSV imagery using a multimodal LLM and a pixel-based semantic-segmentation pipeline, respectively.  Columns compare explanatory power across specifications: GPT4o Only uses the GPT-4o measure alone; ResNet Only uses the ResNet measure alone; Both GPT4o + ResNet includes both image-derived measures; Demographic Only includes the listed covariates; All includes GPT-4o plus demographics.  Estimation is by OLS with robust SEs in parentheses. 
\end{tablenotes}

  \end{threeparttable}
\end{table}

\begin{table}[htbp]
\caption{Model Fit Comparison for GSV--GPT-4o versus GSV-ResNet by Quantile}
\label{tab:Bootstrap_quantile_r2}
\centering
\footnotesize
  \begin{threeparttable}
    \begin{tabular*}{\textwidth}{l @{\extracolsep{\fill}} c c c c c} 
\toprule
& \multicolumn{5}{c}{Quantile}\\
 & 0.10 & 0.25 & 0.50 & 0.75 & 0.90\\
\midrule
\addlinespace[0.3em]
\multicolumn{6}{l}{\textbf{Poverty}}\\
\hspace{1em} GPT-4o & 0.359 & 0.479 & 0.476 & 0.516 & 0.466\\
\hspace{1em} & (0.289--0.427) & (0.434--0.517) & (0.438--0.513) & (0.475--0.557) & (0.396--0.538)\\

\hspace{1em} ResNet & 0.085 & 0.180 & 0.190 & 0.227 & 0.197\\
\hspace{1em} & (0.040--0.144) & (0.145--0.213) & (0.158--0.222) & (0.187--0.268) & (0.138--0.254)\\
\addlinespace[0.3em]
\multicolumn{6}{l}{\textbf{Tree Canopy}}\\
\hspace{1em} GPT-4o & 0.172 & 0.250 & 0.280 & 0.303 & 0.320\\
\hspace{1em} & (0.130--0.232) & (0.197--0.307) & (0.233--0.332) & (0.245--0.359) & (0.247--0.380)\\

\hspace{1em} ResNet & 0.060 & 0.064 & 0.080 & 0.086 & 0.091\\
\hspace{1em} & (0.019--0.102) & (0.036--0.097) & (0.056--0.105) & (0.050--0.123) & (0.034--0.145)\\
\addlinespace[0.3em]
\multicolumn{6}{l}{\textbf{Sustainability Index}}\\
\hspace{1em} GPT-4o & 0.433 & 0.493 & 0.478 & 0.487 & 0.465\\
\hspace{1em} & (0.368--0.488) & (0.449--0.533) & (0.433--0.514) & (0.441--0.526) & (0.401--0.524)\\

\hspace{1em} ResNet & 0.213 & 0.202 & 0.191 & 0.181 & 0.189\\
\hspace{1em} & (0.162--0.252) & (0.162--0.238) & (0.160--0.221) & (0.144--0.219) & (0.141--0.238)\\
\bottomrule
\end{tabular*}
\begin{tablenotes}[flushleft]
\item \scriptsize\textit{Notes:} Entries report pseudo-$R^2$ from $\tau$--quantile regressions of authoritative outcomes (ACS poverty; GEIE canopy; SI $= z(\text{canopy}) - z(\text{poverty})$) on method--specific predictions. Parentheses give percentile 95\% confidence intervals from $B=500$ CBG--level bootstrap resamples. Higher pseudo-$R^2$ indicates closer alignment with the authoritative target at that $\tau$.  See Panel~B of Fig.~\ref{fig:rsquared_analysis}.

\end{tablenotes}

  \end{threeparttable}
\end{table}

\begin{table}[!htbp]
\centering
\caption{Redlining Effects from the Stacked Regression }
\label{tab:All_interactions}
\begin{threeparttable}
\footnotesize
\begin{tabular}{l c c c c c c}
\toprule
& \multicolumn{2}{c}{Poverty} & \multicolumn{2}{c}{Canopy} & \multicolumn{2}{c}{SI} \\
\cmidrule(lr){2-3}\cmidrule(lr){4-5}\cmidrule(lr){6-7}

 & Zipcode FE & SAR & Zipcode FE & SAR& Zipcode FE & SAR\\
\midrule
Redlined (ACS/GEIE) & $1.11^{***}$  & $0.58^{***}$  & $-0.83^{***}$ & $-0.33^{**}$ & $-0.97^{***}$ & $-0.45^{***}$ \\
                              & $(0.20)$      & $(0.09)$      & $(0.25)$      & $(0.11)$     & $(0.18)$      & $(0.08)$      \\
Redlined × GPT-4o             & $-0.11$       & $-0.11$       & $0.35^{*}$    & $0.36^{*}$   & $0.23^{*}$    & $0.24^{*}$    \\
                              & $(0.13)$      & $(0.12)$      & $(0.16)$      & $(0.15)$     & $(0.11)$      & $(0.11)$      \\
Redlined × ResNet             & $-0.78^{***}$ & $-0.79^{***}$ & $0.00$        & $0.01$       & $0.39^{***}$  & $0.40^{***}$  \\
                              & $(0.13)$      & $(0.12)$      & $(0.16)$      & $(0.15)$     & $(0.11)$      & $(0.11)$      \\
\bottomrule
\end{tabular}
\begin{tablenotes}[flushleft]

\item \scriptsize\textit{Notes:} $^{*}$ denotes statistical significance at the 0.05 level. Entries are standardized coefficients (SE in parentheses) from stacked regressions using a common CBG sample.  The baseline ACS/GEIE redlining effect is given in the first row; interaction terms show deviations under GPT-4o and ResNet. The implied total effects for these approaches equal the baseline plus their respective interaction. The corresponding Fig.~\ref{fig:Bootstrap_Interaction_Effects} shows the empirical bootstrap distributions of the total effects by methodological appraoch $m$, as in Eq. \ref{eq:stacked}, providing nonparametric uncertainty bands that accommodate repeated CBGs in the stacked dataset.

\end{tablenotes}
\end{threeparttable}
\end{table}

\end{document}